\documentclass[twocolumn, twocolappendix]{aastex631}

\hypersetup{linkcolor=red,citecolor=blue,filecolor=cyan,urlcolor=blue}
\shorttitle{The Reservoir of the Per-emb-2 Streamer}
\shortauthors{Taniguchi et al.}
\graphicspath{{./}{figures/}}

\begin{document}

\title{The Reservoir of the Per-emb-2 Streamer}

\correspondingauthor{Kotomi Taniguchi}
\email{kotomi.taniguchi@nao.ac.jp}

\author[0000-0003-4402-6475]{Kotomi Taniguchi}
\affiliation{National Astronomical Observatory of Japan, National Institutes of Natural Sciences, 2-21-1 Osawa, Mitaka, Tokyo 181-8588, Japan}

\author[0000-0002-3972-1978]{Jaime E Pineda}
\affiliation{Center for Astrochemical Studies, Max-Planck-Institut f\"{u}r Extraterrestrische Physik, Gie{\ss}enbachstrasse 1, D-85741 Garching bei M\"{u}nchen, Germany}

\author[0000-0003-1481-7911]{Paola Caselli}
\affiliation{Center for Astrochemical Studies, Max-Planck-Institut f\"{u}r Extraterrestrische Physik, Gie{\ss}enbachstrasse 1, D-85741 Garching bei M\"{u}nchen, Germany}

\author[0000-0002-1054-3004]{Tomomi Shimoikura}
\affiliation{Faculty of Social Information Studies, Otsuma Women's University, Sanban-cho, Chiyoda, Tokyo 102-8357, Japan}

\author[0000-0001-7594-8128]{Rachel K. Friesen}
\affiliation{Department of Astronomy \& Astrophysics, University of Toronto, 50 St. George St., Toronto, ON, M5S 3H4, Canada}

\author[0000-0003-3172-6763]{Dominique M. Segura-Cox}\thanks{NSF Astronomy and Astrophysics Postdoctoral Fellow}
\affiliation{Department of Astronomy, The University of Texas at Austin, 2500 Speedway, Austin, TX 78712, USA}
\affiliation{Center for Astrochemical Studies, Max-Planck-Institut f\"{u}r Extraterrestrische Physik, Gie{\ss}enbachstrasse 1, D-85741 Garching bei M\"{u}nchen, Germany}

\author[0000-0002-1730-8832]{Anika Schmiedeke}
\affiliation{Green Bank Observatory, PO Box 2, Green Bank, WV 24944, USA}

\begin{abstract}
Streamers bring gas from outer regions to protostellar systems and could change the chemical composition around protostars and protoplanetary disks.
We have carried out mapping observations of carbon-chain species (HC$_3$N, HC$_5$N, CCH, and CCS) in the 3mm and 7mm bands toward the streamer flowing to the Class 0 young stellar object (YSO) Per-emb-2 with the Nobeyama 45m radio telescope.
A region with a diameter of $\sim0.04$ pc is located north with a distance of $\sim 20,500$ au from the YSO.
The streamer connects to this north region which is the origin of the streamer. 
The reservoir has high density and low temperature ($n_{\rm {H}_2} \approx 1.9 \times 10^4$ cm$^{-3}$, $T_{\rm {kin}} = 10$ K), which are similar to those of early stage starless cores.
By comparisons with the observed abundance ratios of CCS/HC$_3$N to the chemical simulations, the reservoir and streamer are found to be chemically young.
The total mass available for the streamer is derived to be $24-34$ M$_{\odot}$.
If all of the gas in the reservoir will accrete onto the Per-emb-2 protostellar system, the lifetime of the streamer has been estimated at (1.1 -- 3.2)$\times10^{5}$ yr, suggesting that the mass accretion via the streamer would continue until the end of the Class I stage.
\end{abstract}

\keywords{Astrochemistry (75) --- Low mass stars (2050) --- Star formation (1569) --- Young stellar objects (1834)} 

\section{Introduction} \label{sec:intro}

Streamers, velocity coherent structures funneling material around young stellar objects (YSOs), bring gas from outer regions to the disk forming regions \citep[for a review see][]{2023ASPC..534..233P}.
The outer gas would have different chemical compositions from the gas close to YSOs and then these streamers could change the chemical composition at the protoplanetary disks and/or the protostellar cores.
Their kinematic structures are composed of a smooth velocity gradient driven by gravity and rotation.
Streamers may play essential roles during the formation of the solar system \citep{2023ApJ...947L..29A}.
Thus, it is important to reveal the physics and chemistry of these streamer structures to understand the star and planet formation including our solar system.

Many observations with interferometers have revealed the presence of streamers toward various evolutionary stage YSOs; highly embedded Class 0 phase \citep{2019ApJ...885..106L, 2020NatAs...4.1158P}, Class I stage \citep{2016ApJ...823..151C,2022A&A...667A..12V}, and Class II stage \citep{2021ApJ...908L..25G, 2022A&A...658A.104G}.
Similar infall gas structures feeding circumbinary disks around binary systems have also been found \citep{2017ApJ...837...86T, 2019Sci...366...90A}.
In addition, triple spiral arms connected to a triple protostellar system IRAS\,042329+2436 have been found using the SO emission as seen by the Atacama Large Millimeter/submillimeter Array \citep[ALMA,][]{2023ApJ...953...82L}.
They compared the observational results with numerical simulations and suggested that the large triple spiral arms are produced by gravitational interactions between compact triple protostars and the turbulent infalling envelope.
In the high-mass regime, \citet{2019A&A...628A...6S} found converging flows toward the high-mass star-forming cluster Sagittarius B2 (N). 
The protostellar system GGD 27-MM1 has been shown to present streamers \citep{2023ApJ...956...82F}.
Thus, mass accretion via streamers seems to occur in various protostellar systems.

These streamers are traced by several molecular lines: carbon-chain species (e.g., HC$_{3}$N \citep{2020NatAs...4.1158P} and HC$_5$N \citep{2022A&A...658A..53M}), DCN \citep{2023A&A...669A.137H}, CS, HCO$^{+}$ \citep{2022A&A...658A.104G}, H$_2$CO, SO and SO$_2$ \citep{2022A&A...667A..12V}.
Some of the tracers are likely common among protostellar systems, but others may be unique for each source.
The chemical composition differs between streamers, and suitable tracers would be related to the origin of the streamer and/or environment. 

Many numerical simulations could explain the origin of the streamers \citep[for a review see][]{2023ASPC..534..233P}.
\citet{2015MNRAS.446.2776S} showed that the accretion in the vicinity of protostellar disks is highly anisotropic in their simulations including turbulence and magnetic fields.
Gas capture, where sweeping up gas via Bondi-Hoyle accretion occurs, causes the formation of filamentary arms which appear as accretion streamers \citep{2018MNRAS.475.5618B}.
Another possibility is interactions of a star-disk system with an external perturber such as binary components or a stellar fly-by \citep[e.g.,][]{2017A&A...608A.107V, 2020MNRAS.491..504C}.
We need large-scale mapping observations of the protostellar systems covering their streamers to reveal the mechanisms of the streamers.

The Class 0 YSO IRAS\,03292+3039, also known as Per-emb-2, is a protostellar system located in the Perseus star-forming region \citep[$d=300$ pc;][]{2018ApJ...869...83Z}.
No complex organic molecules (COMs) except for CH$_3$OH were detected in the Perseus ALMA Chemistry Survey \citep[PEACHES;][]{2021ApJ...910...20Y}, and this source is likely a chemically-poor source.
Outflows have been detected with CO lines along the northwest (red shifted) - southeast (blue shifted) direction \citep{2019ApJS..245...21S}.

\citet {2020NatAs...4.1158P} discovered a streamer of this protostellar system with observations using the NOrthern Extended Millimeter Array (NOEMA).
The streamer structure can be seen well in the emission lines of carbon-chain species, such as HC$_3$N and CCS.
On the other hand, the emission of N$_2$H$^+$ and its deuterated isotopologue N$_2$D$^+$ is not consistent with the streamer structure.
This suggests that chemically fresh gas is brought to the disk scale via the streamer \citep{1992ApJ...392..551S,1998ApJ...506..743B}.
\citet{2020NatAs...4.1158P} estimated the mass of the streamer and its infall rate at 0.1 M$_{\odot}$ and $10^{-6}$ M$_{\odot}$\,yr$^{-1}$, respectively, although with a poorly constrained abundance.
The starting point of the streamer and its total mass budget remain unrevealed.

In this paper, we present mapping observations of four carbon-chain molecules, HC$_3$N, HC$_5$N, CCH, and CCS, in the 3mm and 7mm bands toward the streamer of the YSO Per-emb-2 with the Nobeyama 45m radio telescope.
We describe the observations and data reduction in Section \ref{sec:nobe}.
We present analyses including available data obtained by the IRAM 30m telescope and 100m Green Bank Telescope (GBT). 
We briefly describe these observations in Sections \ref{sec:IRAM} and \ref{sec:GBT}, respectively.
Maps obtained by the Nobeyama 45m telescope and line analyses are presented in Sections \ref{sec:map} and \ref{sec:ana}, respectively.
We compare the observed CCS/HC$_3$N abundance ratios to the chemical models to constrain the HC$_3$N abundances (Section \ref{sec:dis1}) and derive the total mass of the streamer (Section \ref{sec:dis2}).
We calculate the streamer infall rate and the lifetime of the streamer in Section \ref{sec:dis3}.
Our main conclusions are summarized in Section \ref{sec:con}.

\section{Observations and Data Reduction} \label{sec:obs}

\subsection{Nobeyama 45m radio telescope} \label{sec:nobe}

\begin{deluxetable*}{llcc}
\tablecaption{Information on the targeted lines \label{tab:line}}
\tablewidth{0pt}
\tablehead{
\colhead{Species} & \colhead{Transition} & \colhead{$E_{\rm {up}}/k$} & \colhead{Frequency} \\
\colhead{} & \colhead{} & \colhead{(K)} & \colhead{(GHz)}
}
\startdata
\multicolumn{4}{l}{\bf{Nobeyama 3mm band observations}} \\
HC$_3$N & $J=9-8$ & 19.6 & 81.8814677  \\
CCS & $J_N=7_6-6_5$ & 15.4 & 81.5051700  \\
CCS & $J_N=6_7-5_6$ & 23.3 & 86.1813910 \\
CCH & $N= 1- 0$, $J=\frac{3}{2}-\frac{1}{2}$, $F= 2- 1$ & 4.2 & 87.3169250  \\
\multicolumn{4}{l}{\bf{Nobeyama 7mm band observations}} \\
HC$_3$N & $J=5-4$ & 6.5 & 45.4903138 \\
HC$_5$N & $J=16-15$ & 17.4 & 42.6021529 \\
HC$_5$N & $J=17-16$ & 19.6 & 45.2647199 \\
CCS & $J_N=4_3-3_2$ & 5.4 & 45.3790460 \\
CC$^{34}$S & $J_N=4_3-3_2$ & 5.3 & 44.4975990 \\
$cyclic$-C$_3$H$_2$ & $J_{Ka,Kc}=3_{2,1}- 3_{1,2}$ & 18.2& 44.1047769 \\
SiO & $J=1-0$ & 2.1 & 43.4237600 \\
HNCO &  $2_{0,2}-1_{0,1}$, $F= 2-1$ & 3.2 & 43.9630395 \\
\multicolumn{4}{l}{\bf{IRAM 30m telescope observations}} \\
HC$_3$N & $J=8-7$ & 15.7 & 72.783822 \\
HC$_3$N & $J=10-9$ & 24.0 & 90.979023 \\
\multicolumn{4}{l}{\bf{Green Bank 100m telescope observations}} \\
CCS & $J_N=2_1-1_2$ & 1.6 & 22.3440308 \\ 
\enddata
\tablecomments{Rest frequencies are taken from the Cologne Database for Molecular Spectroscopy \citep[CDMS;][]{2005JMoSt.742..215M,2016JMoSp.327...95E}, except for CC$^{34}$S which is taken from the JPL catalog \citep{1998JQSRT..60..883P}. We summarize only utilized lines in this paper for observations with the IRAM 30m and Green Bank 100m telescopes.} 
\end{deluxetable*}

We carried out observations with the Nobeyama 45m radio telescope in November and December 2022 and from February to April 2023 (Proposal Number: G22006, PI: Kotomi Taniguchi).
FOREST \citep[FOur beam REceiver System on the 45m Telescope;][]{2016SPIE.9914E..1ZM} and Z45 \citep{2015PASJ...67..117N} receivers were used for the 3mm and 7mm observations, respectively.
The beam size and main beam efficiency ($\eta_{\rm {MB}}$) of FOREST are 18\arcsec and 50\%, and those of Z45 are 37\arcsec and 70\%, respectively.
System temperatures ($T_{\rm {sys}}$) were $\sim 170-300$ K at the 3mm band and $\sim 120-200$ K at the 7mm band, depending on the weather condition and elevation.
The chopper-wheel method was applied for the calibration, and typical calibration errors were 10\% \citep{2017ApJ...844...68T}.

Target lines in each frequency band are summarized in Table \ref{tab:line}.
As our target streamer is expected to have chemically young features as mentioned in Section \ref{sec:intro}, we selected carbon-chain species which are known to be abundant in young starless cores \citep{1992ApJ...392..551S,1998ApJ...506..743B}.
In addition, SiO and HNCO, both of which are shock tracers \citep[e.g.,][]{2010A&A...516A..98R, 2017MNRAS.470L..16P}, were observed simultaneously in the 7mm band.

We used the SAM45 FX-type digital correlator \citep{2012PASJ...64...29K}.
The frequency resolutions were 30.52 kHz and 15.26 kHz for the 3mm and 7mm band observations, respectively.
These frequency resolutions correspond to $\sim0.1$ km\,s$^{-1}$ at each frequency band.
We employed the spectral window mode for FOREST to cover four different lines simultaneously.
The bandwidth was 63 MHz for all of the observations.

We employed the on-the-fly (OTF) observing mode \citep{2007A&A...474..679M}.
In this mode, telescopes are driven smoothly and rapidly across a region of sky, while data and antenna position information are recorded continuously.

The map size is $5\arcmin \times 5\arcmin$, corresponding to 90,000 au (or $\sim 0.44$ pc) squares at the source distance (300 pc).
As we mentioned in Section \ref{sec:intro}, the origin of the streamer was not known, but the streamer comes from the north \citep{2020NatAs...4.1158P}. We then set the map regions covering mainly the northern part of the YSO.
The separations between rows\footnote{\url{https://www.nro.nao.ac.jp/~nro45mrt/html/obs/otf/index_en.html}} were set at 6\arcsec\, and 10\arcsec\, for the 3mm and 7mm band observations. 

The telescope pointing was checked every $1.5-2$ hour by observing the SiO ($J =1-0$) maser line from NML-Tau at ($\alpha_{\rm {J2000}}$, $\delta_{\rm {J2000}}$) = (3$^{\rm {h}}$53$^{\rm {m}}$28\fs86, +11\degr24\arcmin22\farcs4). 
We used the Z45 receiver for the pointing observations.
The pointing error was within 3\arcsec.

We conducted data reduction with the NOSTAR software\footnote{\url{https://www.nro.nao.ac.jp/~nro45mrt/html/obs/otf/reduction_en.html}} 
provided by the Nobeyama Radio Observatory and made fits files for each line.
The pixel size of the maps in the 3mm band is 8\arcsec and that in the 7mm band is 12\arcsec.
Baselines were calculated in line-free channels and the 1st-order polynomial fitting was applied. 
The velocity resolutions are 0.11 km\,s$^{-1}$ for the 3mm band data and 0.15 km\,s$^{-1}$ for the 7mm band data.
The applied convolution function was the Bessel-Gauss function.
We treated the fits files generated by NOSTAR in the Common Astronomy Software Applications (CASA) package \citep{2022PASP..134k4501C} and made maps presented in Section \ref{sec:res}.

\subsection{IRAM 30m Telescope} \label{sec:IRAM}
The observations were carried out with the IRAM 30m telescope at Pico Veleta (Spain) on 2020 May 14 and 15 under project 230-19.
The EMIR E090 receiver and the FTS50 backend were employed. 
We used two spectral setups to cover the HC$_3$N (10--9) and (8--7) lines.
The beam sizes were approximately 33\arcsec and 28\arcsec at 72 GHz and 90 GHz, respectively.
We mapped a region of $\approx$150\arcsec$\times$200\arcsec with the OTF mode \citep{2007A&A...474..679M}. 
The calibration error was 10\%\footnote{\url{https://www.iram.fr/GENERAL/calls/s23/30mCapabilities.pdf}}.
Data reduction was performed using the CLASS program of the GILDAS package\footnote{\url{http://www.iram.fr/IRAMFR/GILDAS}}. 
The beam efficiency, $B_{eff}$, is obtained using the Ruze formula 
(available in CLASS), and it is used to convert the observations 
into main beam temperatures, $T_{\rm mb}$.

\subsection{Green Bank Telescope} \label{sec:GBT}

Observations for the Green Bank Ammonia Survey (GAS) were performed from January 2015 through March 2016 at the Robert C. Byrd Green Bank Telescope (GBT) using the 7-pixel K-band Focal Plane Array (KFPA) and the VErsatile GBT Astronomical Spectrometer (VEGAS). Observational details were first presented in GAS DR1 \citep{2017ApJ...843...63F}, which focused on the NH$_3$ (1,1) and (2,2) emission from four regions. We refer the reader to GAS DR1 for a full description of the observations, data reduction, and imaging of the survey data. 

The VEGAS backend was used in its configuration Mode 20, allowing eight spectral windows per KFPA beam, each with a bandwidth of 23.44~MHz and 4096 spectral channels. The resulting spectral resolution of 5.7~kHz gives a velocity resolution of $\sim 0.07$ km\,s$^{-1}$ at 23.7~GHz. The GAS setup used six of eight available spectral windows to target six spectral lines in each beam, including HC$_3$N (5--4), and CCS ($2_1-1_0$) transition was observed in a single, central beam. In-band frequency switching with a frequency throw of 4.11~MHz was used, maximizing the observing time spent on-source. 

Observations were done in OTF mode \citep{2007A&A...474..679M}, most often scanning in Right Ascension (R.A.) or Declination (Decl.) over square regions of size $10\arcmin \times 10\arcmin$. 
The beam size was 32\arcsec. 
The observed coverage of most regions consists of multiple such observing blocks to cover the desired area, generally observed on different dates. Most $10\arcmin \times 10\arcmin$ blocks were observed once to reach the survey sensitivity goals, but several were observed twice to mitigate the effects of poor weather in the first observations. Pointing updates were usually performed in between completed maps, or more frequently if winds were high ($> 5$ m\,s$^{-1}$). 
The calibration error was 10\% \citep{2017ApJ...843...63F}.

\section{Results} \label{sec:res}

All of the target lines in the 3mm band (see Table \ref{tab:line}) have been detected.
The CCS ($J_N=7_6-6_5$) shows low peak intensity and thus low signal-to-noise (S/N) ratios.
We use another transition ($J_N=6_7-5_6$) and exclude the CCS ($J_N=7_6-6_5$) line from the analyses in the following sections.

In the 7mm band observations, four lines of the carbon-chain species (HC$_3$N, two lines of HC$_5$N, and CCS) have been detected, whereas CC$^{34}$S, $cyclic$-C$_3$H$_2$, SiO, and HNCO were not detected. 
The non-detection of shock tracers (SiO and HNCO) may be caused by the beam dilution effect due to the low angular resolution.
We exclude these undetected lines from the following sections.

\subsection{Maps of carbon-chain species} \label{sec:map}

\begin{figure*}
 \begin{center}
  \includegraphics[bb = 0 20 460 400, scale =0.98]{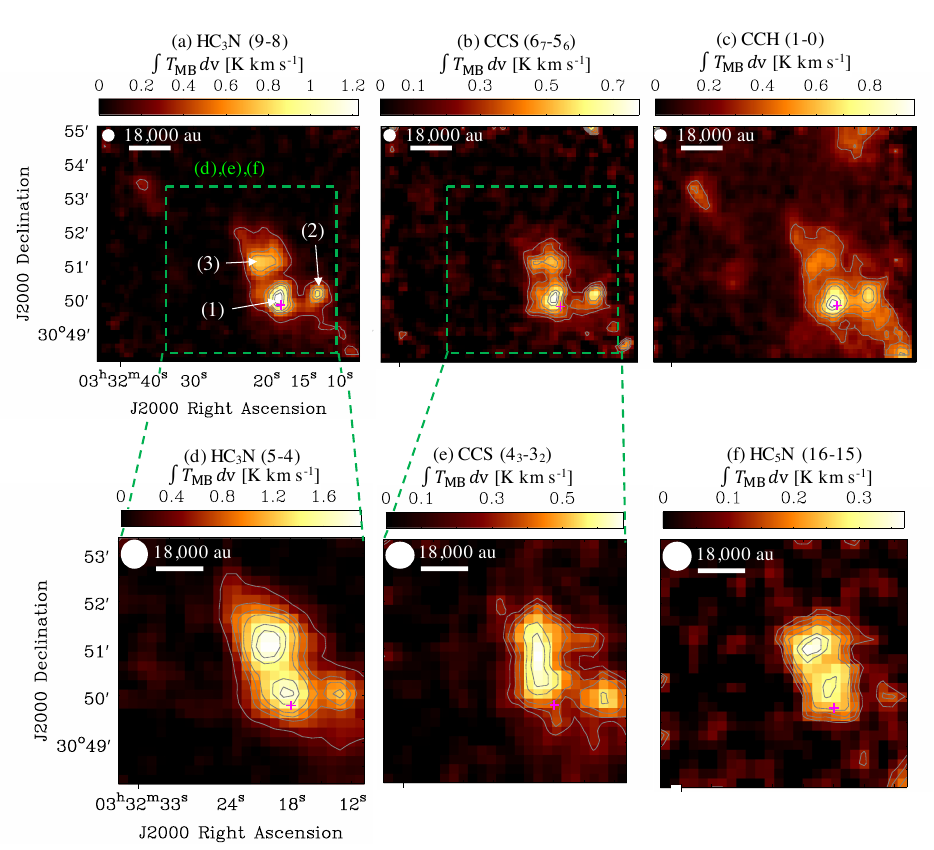}
 \end{center}
\caption{Integrated-intensity maps of (a) HC$_3$N ($9-8$), (b) CCS ($6_7-5_6$), and (c) CCH ($N= 1-0$, $J=\frac{3}{2}-\frac{1}{2}$, $F= 2- 1$), (d) HC$_3$N ($5-4$), (e) CCS ($4_3-3_2$), and (f) HC$_5$N ($16-15$). Contour levels are relative values to their peak levels; (a, d) 20, 40, 50, 60, 80, 90 \%, (b, c, f) 30, 40, 50, 60, 80, 90 \%, (e) 40, 50, 60, 80, 90 \%. The peak levels are (a) 1.27 K km\,s$^{-1}$, (b) 0.78 K km\,s$^{-1}$, (c) 0.97 K km\,s$^{-1}$, (d) 1.90 K km\,s$^{-1}$, (e) 0.68 K km\,s$^{-1}$, and (f) 0.37 K km\,s$^{-1}$, respectively. The integrated velocity ranges are (a) 6.1 -- 8.2 km\,s$^{-1}$, (b) 5.9 -- 7.6 km\,s$^{-1}$, (c) 5.6 -- 8.4 km\,s$^{-1}$, (d) 5.8 -- 8.4 km\,s$^{-1}$, (e) 6.1 -- 8.1 km\,s$^{-1}$, and (f) 6.1 -- 7.6 km\,s$^{-1}$ for each line. The noise levels of these maps are (a) 0.03 K km\,s$^{-1}$, (b) 0.047 K km\,s$^{-1}$, (c) 0.04 K km\,s$^{-1}$, (d) 0.03 K km\,s$^{-1}$,  (e) 0.025 K km\,s$^{-1}$, and (f) 0.05 K km\,s$^{-1}$, respectively. The green dashed squares in panels (a) and (b) indicate the map regions of panels (d) -- (f). Three identified structures are labeled as (1) -- (3) in panel (a). The white-filled circles at the top left corner indicate the beam size (18\arcsec and 37\arcsec at the 3mm and 7mm, respectively). The magenta cross indicates the position of the YSO.\label{fig:mom0}}
\end{figure*}

Upper panels of Figure \ref{fig:mom0} show integrated-intensity maps of (a) HC$_3$N ($J=9-8$), (b) CCS ($J_N=6_7-5_6$), and (c) CCH ($N= 1- 0$, $J=\frac{3}{2}-\frac{1}{2}$, $F= 2- 1$) lines in the 3mm band, and lower panels show the integrated-intensity maps of (d) HC$_3$N ($J=5-4$), (e) CCS ($J_N=4_3-3_2$), and (f) HC$_5$N ($J=16-15$) in the 7mm band.
The HC$_5$N ($J=17-16$) line shows a similar structure to the HC$_5$N ($16-15$) line but shows weak peak intensities and low S/N ratios.
We then only show the result of the HC$_5$N ($16-15$) line.
We present channel maps of the three lines in the 3mm band in Figures \ref{fig:HC3Nchan} -- \ref{fig:CCHchan} in Appendix \ref{sec:a0}.

In panels (a) -- (c) of Figure \ref{fig:mom0}, we can find three structures which are labeled as (1) -- (3) in panel (a); (1) associated with the YSO, (2) located at the west of the YSO, and (3) located north of the first one.
The first structure corresponds to the streamer identified by \citet{2020NatAs...4.1158P}.
The streamer connects to the north structure, and the velocity fields show connections between the YSO and the north structure (see Figures \ref{fig:HC3Nchan} -- \ref{fig:CCHchan} in Appendix \ref{sec:a0}).
Thus, this north structure indicated as (3) in panel (a) is the reservoir of the streamer.
The diameter of the reservoir is approximately 0.04 pc, which is consistent with a typical size of starless cores \citep[e.g.,][]{2018ApJ...853....5P}.
The distance between the YSO and the reservoir is $\sim20,500$ au.
This reservoir is out of the Field of View (FoV) of the NOEMA observations \citep{2020NatAs...4.1158P}.

The three structures identified in the 3mm band data can be recognized in panels (d) HC$_3$N ($5-4$) and (f) HC$_5$N ($16-15$).
On the other hand, the integrated-intensity map of the CCS ($4_3-3_2$) line shows a single peak between the streamer and reservoir.
The west core is spatially resolved.
This is probably caused by the larger beam size at 45 GHz (37\arcsec) compared to 80 GHz (18\arcsec).

\subsection{Spectral analyses} \label{sec:ana}

\begin{figure*}
 \begin{center}
  \includegraphics[bb = 0 20 520 250, scale =0.7]{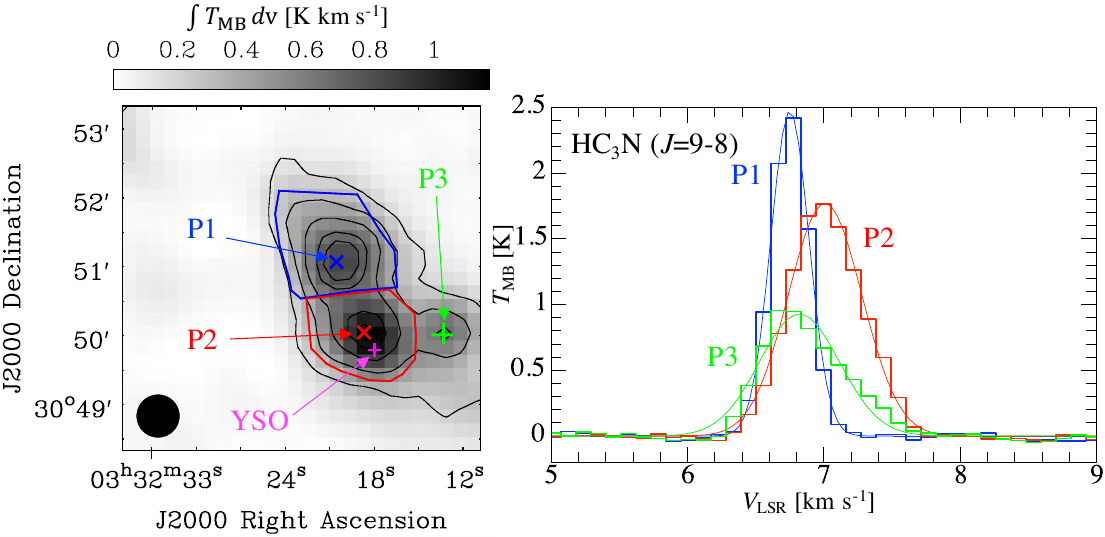}
 \end{center}
\caption{Left panel shows a regridded integrated-intensity map of HC$_3$N ($J=9-8$) overlaid with contours indicating the HC$_3$N ($J=5-4$) line (contour levels are 20, 40, 60, 80, 90\% of the peak intensity). Two crosses (P1 and P2) indicate two positions where we conducted spectral analyses. The polygons harboring P1 and P2 indicate the regions where we derived the mass. The green cross indicated as P3 is the other emission peak. The magenta cross indicates the position of the YSO. Right panel shows spectra of the HC$_3$N ($J=9-8$) line at P1 (blue), P2 (red), and P3 (green) respectively. The curves show fitting results with the Gaussian profile. The fitting parameters are summarized in Table \ref{tab:HC3Nline} in Appendix \ref{sec:a1}. \label{fig:HC3Nregrid}}
\end{figure*}

We smoothed and regridded all maps to have the same beam grids as the 7mm band data obtained with the Nobeyama 45m telescope (37\arcsec) .
The left panel of Figure \ref{fig:HC3Nregrid} shows a regridded integrated-intensity map of the HC$_3$N ($J=9-8$) line (grayscale) overlaid with the integrated-intensity map of its $J=5-4$ line (black contours).
Two emission peaks, corresponding to the streamer and the reservoir, are well consistent with the two lines. 

We derived column densities of HC$_3$N, CCS, and HC$_5$N at the two peaks indicated as P1 and P2 in the left panel of Figure \ref{fig:HC3Nregrid}.
The coordinates of YSO and the HC$_3$N emission peaks (P1 -- P3) are summarized in Table \ref{tab:coordinate}.

\begin{deluxetable}{lcc}
\tablecaption{The coordinates of Per-emb-2 YSO and emission peaks of HC$_3$N \label{tab:coordinate}}
\tablewidth{0pt}
\tablehead{
\colhead{Position} & \colhead{R.A.(J2000)} & \colhead{Decl.(J2000)} 
}
\startdata
YSO & 3$^{\rm {h}}$32$^{\rm {m}}$17\fs96 & +30\degr49\arcmin47\farcs5 \\
P1 & 3$^{\rm {h}}$32$^{\rm {m}}$20\fs53 & +30\degr51\arcmin04\farcs5 \\
P2 & 3$^{\rm {h}}$32$^{\rm {m}}$18\fs68 & +30\degr50\arcmin03\farcs2 \\
P3 & 3$^{\rm {h}}$32$^{\rm {m}}$13\fs16 & +30\degr50\arcmin02\farcs6 \\
\enddata
\end{deluxetable}

\begin{deluxetable*}{lcccccc}
\tablecaption{Results of line analyses \label{tab:lineres}}
\tablewidth{0pt}
\tablehead{
\colhead{} & \colhead{} &  \multicolumn{2}{c}{P1} & & \multicolumn{2}{c}{P2} \\
\cline{3-4} \cline{6-7}
\colhead{Species} & \colhead{Analytical Method} & \colhead{$N$ (cm$^{-2}$)} & \colhead{$T_{\rm {ex}}$ (K)} & \colhead{} & \colhead{$N$ (cm$^{-2}$)} & \colhead{$T_{\rm {ex}}$ (K)}
}
\startdata
HC$_3$N  & SLED & ($2.6\pm0.2$)$\times 10^{13}$ & ... & & ($9.4\pm0.2$)$\times 10^{12}$ & ...  \\
CCS & Rotational Diagram &  ($8.5\pm1.9$)$\times 10^{12}$ & $10.5\pm1.8$ & & ($1.1\pm0.3$)$\times 10^{13}$ & $13.7\pm3.7$  \\
HC$_5$N & MCMC & ($8.9\pm0.9$)$\times 10^{12}$ & $10.7\pm0.7$ & & ($9.5\pm0.9$)$\times10^{12}$ & $13.03 \pm 0.03$ \\
\enddata
\tablecomments{The errors indicate the standard deviation.} 
\end{deluxetable*}

\subsubsection{Spectral-line energy distribution of HC$_3$N} \label{sec:HC3NSLED}

\begin{figure*}
 \begin{center}
  \includegraphics[bb = 0 30 530 190, scale =0.8]{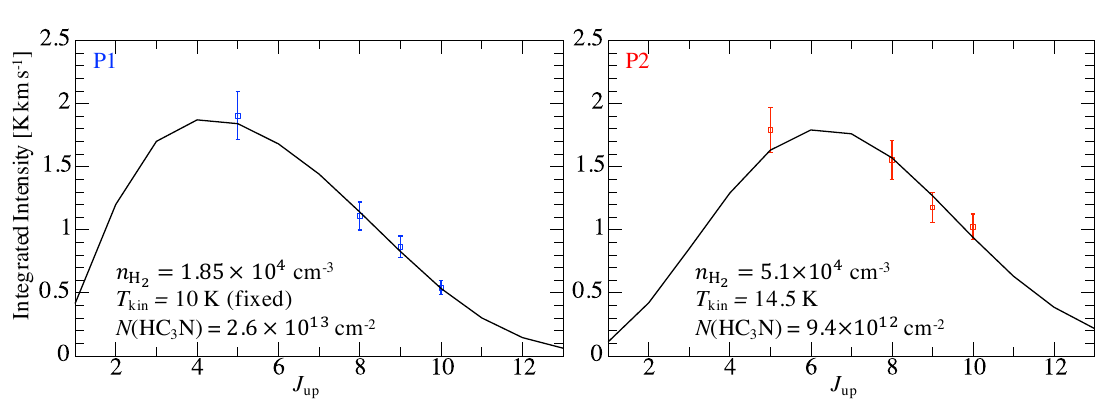}
 \end{center}
\caption{Results of the SLED analysis with HC$_3$N lines toward P1 (left) and P2 (right). The blue and red points indicate the observed values. The 10\% errors due to a typical uncertainty of the chopper-wheel method are taken into consideration. The black curves indicate the best fitting result with the parameters shown in each panel; $n_{\rm {H_{2}}}= 1.85\times10^4$ cm$^{-3}$, $T_{\rm {kin}}=10$ K (fixed) and $N$(HC$_3$N)$=2.6\times10^{13}$ cm$^{-2}$ for P1, and  $n_{\rm {H_{2}}}= 5.1\times10^4$ cm$^{-3}$, $T_{\rm {kin}}=14.5$ K, and $N$(HC$_3$N)$=9.4\times10^{12}$ cm$^{-2}$ for P2, respectively.  \label{fig:SLED}}
\end{figure*}

We conducted the spectral-line energy distribution (SLED) analysis for the HC$_3$N data with the non-LTE code RADEX \citep{2007A&A...468..627V}.
The collisional parameters were obtained from \citet{2016MNRAS.460.2103F} which distinguish between $ortho$-H$_2$ and $para$-H$_2$.
We searched for the best combination parameter sets to reproduce the observed integrated intensities with the least chi-square method by changing the H$_2$ density ($n_{\rm {H_{2}}}$) and column density of HC$_3$N.

The line widths at each position were derived from fitting the HC$_3$N ($9-8$) spectra with the Gaussian profile; 0.33 km\,s$^{-1}$ and 0.65 km\,s$^{-1}$ for P1 and P2, respectively (see the right panel of Figure \ref{fig:HC3Nregrid}).
Table \ref{tab:HC3Nline} in Appendix \ref{sec:a1} summarizes the line parameters obtained by fitting with the Gaussian profile at P1 -- P3.
The line widths are well consistent with the results obtained with NOEMA \citep[see Figure 2 (b) in][]{2020NatAs...4.1158P}\footnote{The relationship between FWHM and $\sigma$ is as follows: FWHM = 2$\sigma\sqrt{2{\rm {ln}}2}$.}.
The wider line width at P2 seems to reflect a larger velocity gradient and/or accretion shock.
The velocity component at P2 is well consistent with the systemic velocity (7.05 km\,s$^{-1}$) reported by \citet{2020NatAs...4.1158P}.

The constraints of the gas kinetic temperatures were done by the NH$_3$ data taken with the GBT.
We fixed the gas kinetic temperature ($T_{\rm {kin}}$) at 10 K for P1. 
In the case of P2, we changed $T_{\rm {kin}}$ from 13 K to 15 K with the 0.5-K step and determined it with $n_{\rm {H_{2}}}$ and $N$(HC$_3$N) simultaneously.
The background temperature is fixed at 2.73 K, the standard cosmic microwave background temperature.
The tested ranges of H$_2$ density and HC$_3$N column density were $5\times10^3 - 5\times10^5$ cm$^{-3}$ (step is log($n$) = 0.073) and $5\times10^{12} - 5\times10^{13}$ cm$^{-2}$ (step is $1.8\times10^{11}$ cm$^{-2}$), respectively.

The results of the SLED analyses with the best parameter combinations are presented in Figure \ref{fig:SLED}.
The values at P1 are derived to be $n_{\rm {H_{2}}}=$($1.9\pm0.3$)$\times10^4$ cm$^{-3}$ and $N$(HC$_3$N)=($2.6\pm0.2$)$\times10^{13}$ cm$^{-2}$.
For the P2 position, we determined the values of $n_{\rm {H_{2}}}$, $T_{\rm {kin}}$, and $N$(HC$_3$N) at ($5.1\pm0.6$)$\times10^4$ cm$^{-3}$, $14.5 \pm 0.3$ K, and ($9.4\pm0.2$)$\times10^{12}$ cm$^{-2}$, respectively.
Table \ref{tab:lineres} summarizes the obtained column densities of HC$_3$N at the two positions.

\subsubsection{Rotational Diagram of CCS} \label{sec:CCSRD}

\begin{figure*}
 \begin{center}
  \includegraphics[bb = 0 15 500 455, scale =0.75]{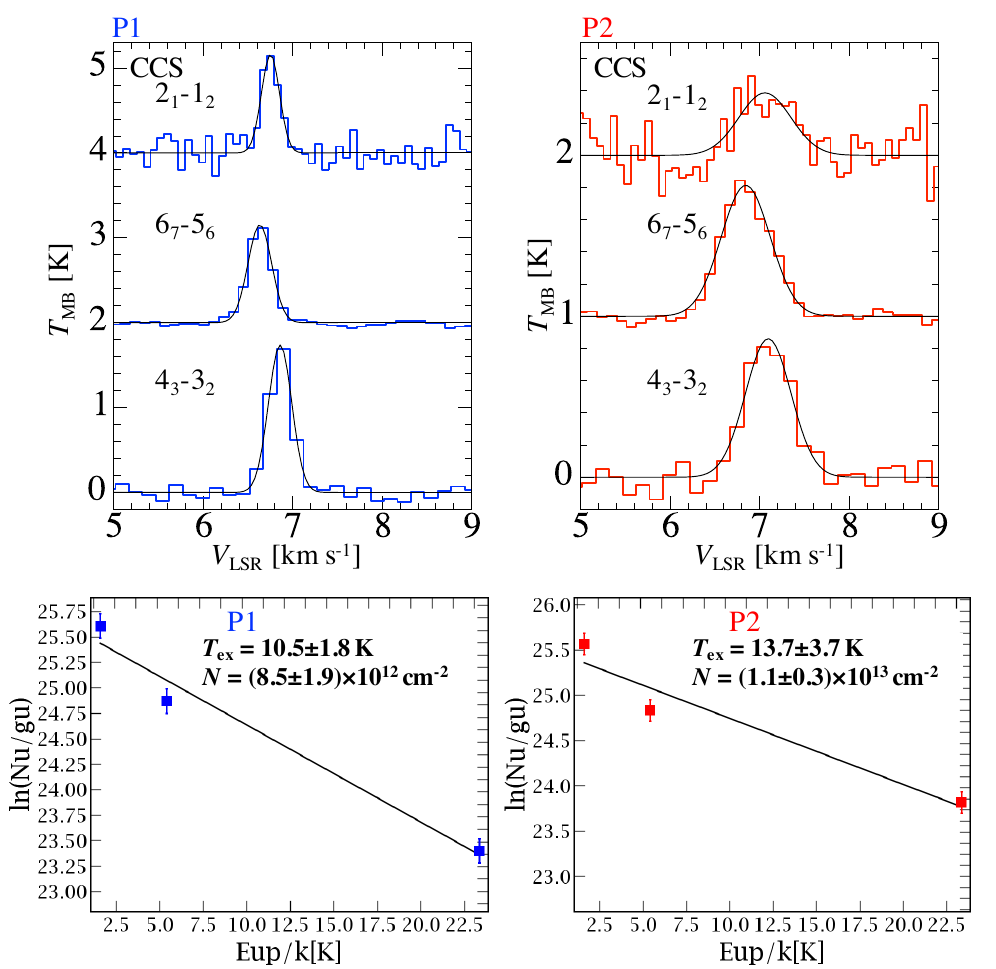}
 \end{center}
\caption{Upper panels show CCS spectra of the three transitions at P1 (left) and P2 (right). Blue and red lines indicate the observed spectra, and black curves indicate the results of the Gaussian fitting. Lower panels show rotational diagrams of the CCS lines at P1 (left) and P2 (right), respectively. 
Errors are calculated from the Gaussian fitting errors and the calibration errors (10\%). \label{fig:RD}}
\end{figure*}

Since there are no available collisional parameters of CCS, the SLED analysis could not be applied to the CCS lines.
We have conducted the rotational diagram analysis for the CCS data.
We fitted the spectra with the Gaussian profile and conducted the rotational diagram analysis in the CASSIS software \citep{2015sf2a.conf..313V}.

The upper panels of Figure \ref{fig:RD} show the spectra of the three CCS lines at P1 (left) and P2 (right).
Black curves indicate the results of the Gaussian fitting. 
The derived parameters are summarized in Table \ref{tab:CCSline} in Appendix \ref{sec:a1}.
Line widths (FWHM) are consistent with those of HC$_3$N at each position.
The $6_7-5_6$ transition lines have peaks at marginally lower velocities compared to the other two lines by $\sim0.15$ km\,s$^{-1}$, which may suggest that this line traces different regions due to its higher upper-state energy.
However, we need high-velocity resolution data to confirm it, because our data have velocity resolutions around 0.11 -- 0.15 km\,s$^{-1}$. 

The lower panels of Figure  \ref{fig:RD} show the rotational diagrams at P1 (left) and P2 (right), respectively.
The obtained excitation temperatures at P1 and P2, $10.5\pm1.8$ K and $13.7\pm3.7$ K, which agree with the gas kinetic temperatures derived by the SLED analysis using the HC$_3$N data (Section \ref{sec:HC3NSLED}) within their errors.
The column density at P2 is derived to be ($1.1\pm0.3$)$\times10^{13}$ cm$^{-2}$, which is slightly higher compared to P1 (($8.5\pm1.9$)$\times10^{12}$ cm$^{-2}$).
Table \ref{tab:lineres} summarizes these results.
We checked the optical depth ($\tau$) of these lines in the CASSIS software by synthesizing spectra with the obtained column densities, rotational temperatures, and line widths.
We confirmed that all of the lines are optically thin ($\tau \approx 0.11 - 0.4$). 
Thus, the rotational diagram analysis is reasonable.
An optical depth of 0.4 may cause an underestimation of the column density at most 20\%, which is smaller than the fitting errors ($\sim22-27$\%). Hence, we conclude that the optical thickness does not affect our results significantly.

\subsubsection{Markov Chain Monte Carlo of HC$_5$N} \label{sec:HC5NMCMC}

\begin{figure}
 \begin{center}
  \includegraphics[bb = 0 25 500 270, scale =0.6]{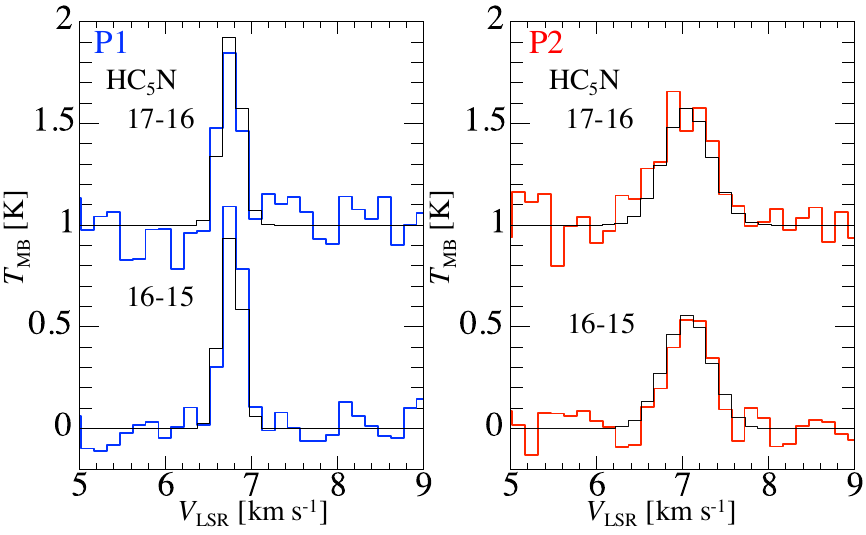}
 \end{center}
\caption{Two HC$_5$N spectra at P1 (left) and P2 (right) overlaid by the best fitting results obtained by the MCMC method (black lines). \label{fig:MCMC}}
\end{figure}

As we have two HC$_5$N lines with similar upper-state energies, the rotational diagram method could not be applied.
We analyzed the HC$_5$N spectra with the Markov Chain Monte Carlo (MCMC) method assuming the local thermodynamic equilibrium (LTE) condition in the CASSIS software \citep{2015sf2a.conf..313V}.
The critical density of the HC$_5$N line is $\sim10^4$ cm$^{-3}$ \citep{2009ApJ...699..585H,2016ApJ...817..147T}, which is similar to the H$_2$ densities at P1 and P2, and then the LTE assumption is reasonable here.

The column density ($N$), excitation temperature ($T_{\rm {ex}}$), velocity component of the line peak ($V_{\rm {LSR}}$), and line width (FWHM) were treated as free parameters.
We constrained the excitation temperatures at 9.5 -- 12.0 K and 10.0 -- 15.0 K at P1 and P2, respectively (see Sections \ref{sec:HC3NSLED} and \ref{sec:CCSRD}).
The tested ranges of FWHM and $V_{\rm {LSR}}$ were 0.1 -- 0.8 km\,s$^{-1}$ and 6.0 -- 7.5 km\,s$^{-1}$ at P1, and 0.1 -- 1.6 km\,s$^{-1}$ and 6.4 -- 7.5 km\,s$^{-1}$ at P2.
The range of the column density was $2.0 \times 10^{12}$ -- $8.0 \times 10^{13}$ cm$^{-2}$ at both P1 and P2.
Since the emission size is much larger than the beam size (37\arcsec), the beam dilution effect does not affect the analysis.

Blue and red lines in Figure \ref{fig:MCMC} show the observed spectra of two HC$_5$N lines at P1 (left) and P2 (right).
The black lines indicate the best-fitting model for each spectrum. 
The optical depths are $\sim 0.30-0.33$ and $\sim0.093-0.095$ at P1 and P2, respectively.
Thus, the assumption of the LTE is reasonable.

Table \ref{tab:lineres} summarizes derived column densities and excitation temperatures at each position.
The derived FWHM and $V_{\rm {LSR}}$ are $0.28\pm0.02$ km\,s$^{-1}$ and $6.765\pm0.008$ km\,s$^{-1}$ at P1, and $0.63\pm0.03$ km\,s$^{-1}$ and $7.06\pm0.02$ km\,s$^{-1}$ at P2, respectively.
These line widths (FWHM) are consistent with those of HC$_3$N and CCS, and the velocity components match with those of HC$_3$N at each position (see Tables \ref{tab:HC3Nline} and \ref{tab:CCSline} in Appendix \ref{sec:a1}).

\section{Discussion} \label{sec:dis}

We derive masses of the reservoir (P1) and the streamer (P2).
Blue and red polygons in the left panel of Figure \ref{fig:HC3Nregrid} indicate regions where masses are derived.
Hereafter, we call them ``Region P1'' and ``Region P2'', respectively.

\subsection{Comparisons with chemical simulation} \label{sec:dis1}

\begin{figure}
 \begin{center}
  \includegraphics[bb = 0 35 480 460, scale =0.5]{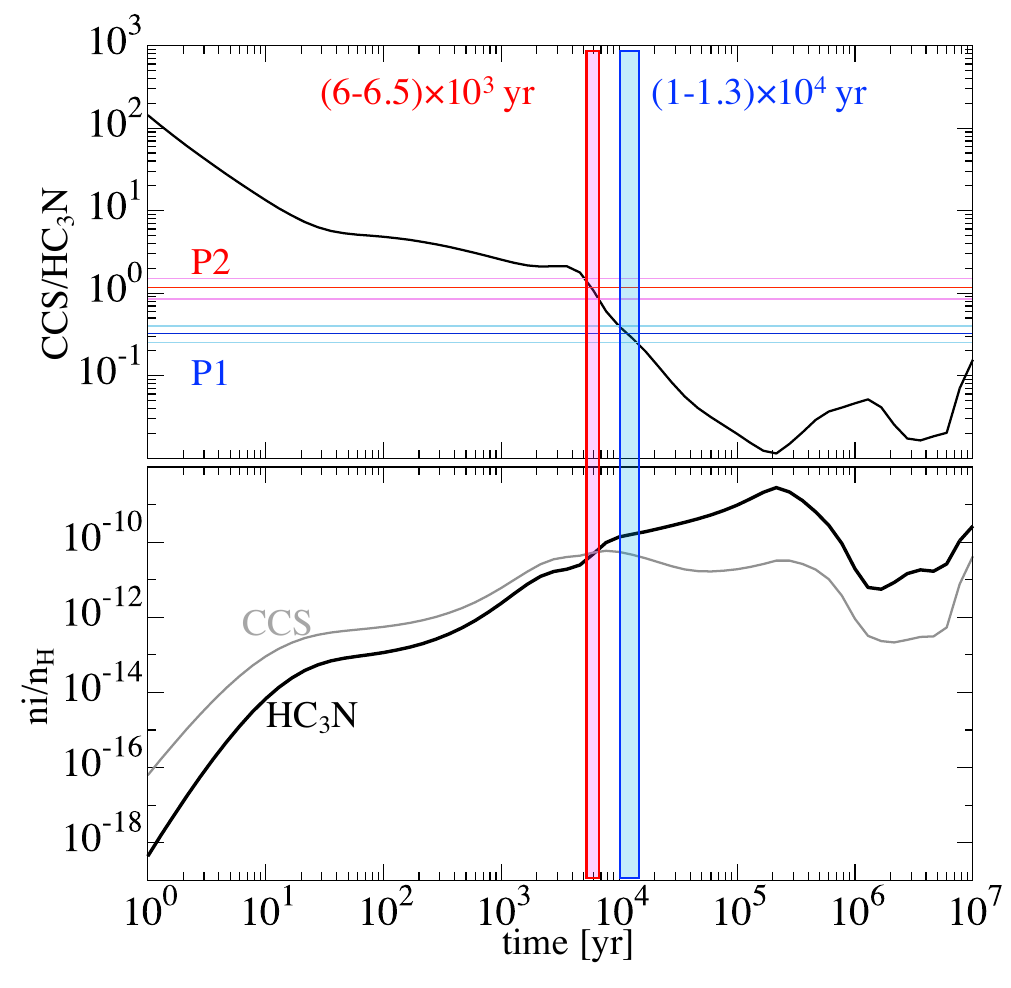}
 \end{center}
\caption{The upper panel shows comparisons of the observed CCS/HC$_3$N abundance ratios at P1 and P2 to the modeled results. The horizontal lines show the observed values (light-colored lines indicate the upper and lower limits). The lower panel shows abundances of HC$_3$N (black) and CCS (gray). The shaded vertical regions indicate the best-fit ages. \label{fig:modelHC3N}}
\end{figure} 

We ran chemical simulations with a constant temperature and density model utilized for starless cores \citep{2019ApJ...884..167T} to constrain the HC$_3$N abundances from the CCS/HC$_3$N abundance ratios.
We used the gas-grain Nautilius code \citep{2016MNRAS.459.3756R}.
The temperature and density are fixed at 10 K and $10^4$ cm$^{-3}$, respectively.
This condition is constrained by the results of the SLED analyses (Section \ref{sec:HC3NSLED}).
The P1 position corresponds to the emission peak of carbon-chain species, and the density at P1 is likely higher compared to the environments.
Since the density and temperature do not change drastically between P1 and P2, this assumption is reasonable for most of the gas within these regions.

We applied the three-phase model, in which the chemistry of the gas phase, the grain surface, and the bulk ice are treated.
We applied the cosmic-ray ionization rate of $1.3\times10^{-17}$ s$^{-1}$.
All of the parameters are the same as in the chemical code described in \citet{2019ApJ...884..167T}, which is used for the chemical simulations of starless cores in nearby low-mass star-forming regions.

We compared the CCS/HC$_3$N abundance ratios at P1 and P2 to the chemical model (upper panel of Figure \ref{fig:modelHC3N}).
The observed CCS/HC$_3$N abundance ratios at P1 and P2 are $0.33\pm0.07$ and $1.2\pm0.3$, respectively.
Table \ref{tab:modelcomp} summarizes the best-fit ages and the HC$_3$N abundances at the best-fit ages (see the lower panel of Figure \ref{fig:modelHC3N}) at P1 and P2.
The best-fit ages are (1 -- 1.3)$\times10^4$ yr and (6 -- 6.5)$\times10^3$ yr at P1 and P2, respectively.
The different ages at P1 and P2 positions may be caused by different emission regions for HC$_3$N and CCS.
Since the critical densities of the HC$_3$N ($5-4$) and CCS ($4_3-3_2$) lines are more or less similar  \citep[$n_{{\rm {H}}_2} \approx 10^{4-5}$ cm$^{-3}$;][]{2018ApJ...864...82D}, these two lines are expected to trace similar regions. 
Here, we simply assume that both lines trace the same regions.
However, integrated-intensity maps of these two lines in the 7mm band (panels (d) and (e) of Figure \ref{fig:mom0}) show different features.
This may be caused by the low angular resolution, but we could not completely exclude the possibility that these lines trace slightly different regions.

We also conducted the same analysis using the CCS/HC$_5$N abundance ratios as shown in the upper panel of Figure \ref{fig:modelHC5N} in Appendix \ref{sec:a2}.
The best-fit ages are around (2.3 -- 3.0)$\times10^4$ yr at P1 and P2.
Thus, P1 and P2 have chemically young characteristics such that carbon-chain species are abundant (see the lower panel of Figure \ref{fig:modelHC5N} in Appendix \ref{sec:a2}).

\citet{2020NatAs...4.1158P} found that the emission of N$_2$H$^+$ and its deuterium species do not trace the streamer.
This is explained by the fact that the streamer brings in chemically fresh gas.
We checked this point with the results of the chemical model.
The lower panel of Figure \ref{fig:modelHC5N} in Appendix \ref{sec:a2} shows the time evolution of abundances of the three carbon-chain species (CCS, HC$_3$N, HC$_5$N) and N$_2$H$^+$.
Around $10^4$ yr, the abundance of N$_2$H$^+$ ($\sim10^{-13}$) is lower than those of the carbon-chain species by more than two orders of magnitude.
Thus, the streamer cannot be traced by the N$_2$H$^+$ lines as the observations showed \citep{2020NatAs...4.1158P}.

All of the results agree that the reservoir and the streamer have chemically young characteristics, and their chemical ages are very similar.
This is other evidence that Region P1 is the origin of the streamer, $i.e.,$ the reservoir.
 
\subsection{Masses of the reservoir and the streamer} \label{sec:dis2}

Using the HC$_3$N column densities (see Table \ref{tab:lineres} and Section \ref{sec:HC3NSLED}) and its constrained abundances at P1 and P2 (Table \ref{tab:modelcomp}), we derived the H$_2$ column densities, denoted as $N$(H$_2$), at P1 and P2\footnote{$N$(H$_2$) = $N$(HC$_3$N)/2$X$(HC$_3$N)}.
The $N$(H$_2$) values are derived to be ($7.7-9.5$)$\times10^{22}$ cm$^{-2}$ and ($6.7-11.8$)$\times10^{22}$ cm$^{-2}$ at P1 and P2, respectively (Table \ref{tab:modelcomp}).

We derived masses of Region P1 and Region P2 within the blue and red polygons in the left panel of Figure \ref{fig:HC3Nregrid}.
The masses within Region P1 (the reservoir; $M_{\rm {Region 1}}$) and Region P2 (the streamer; $M_{\rm {Region 2}}$) are derived to be 14.5 -- 17.9 M$_{\odot}$ and 9.3 -- 16.2 M$_{\odot}$, respectively.
Thus, the total mass available for the streamer is 24 -- 34 M$_{\odot}$.

\citet{2020NatAs...4.1158P} derived the mass of the streamer to be 0.1 M$_{\odot}$, and their results underestimated the mass of the streamer.
This is caused by the fact that they assumed a very high HC$_3$N abundance ([HC$_3$N]/[H$_2$] = $7.4 \times 10^{-9}$), which is comparable to that at the Cyanopolyyne Peak in Taurus Molecular Cloud-1 where carbon-chain molecules are extraordinarily abundant \citep[e.g.,][]{1992ApJ...392..551S,2004PASJ...56...69K,2018ApJ...864...82D}.
Their applied value is higher than those we derived by two orders of magnitude.
In addition, we calculated the masses within larger areas compared to \citet{2020NatAs...4.1158P}.
These differences cause the different derived masses between the two studies.
The reservoir (Region P1), which was not covered by the previous NOEMA observations \citep{2020NatAs...4.1158P}, harbors gas that is comparable to or more abundant than the streamer.

\begin{deluxetable}{lcc}
\tablecaption{Summary of results from comparisons with chemical simulations \label{tab:modelcomp}}
\tablewidth{0pt}
\tablehead{
\colhead{Parameters} & \colhead{P1} & \colhead{P2}
}
\startdata
Age [yr] & ($1-1.3$)$\times10^4$ & ($5-6.5$)$\times10^3$ \\
$X$(HC$_3$N) & ($1.38-1.7$)$\times10^{-10}$ & ($3-7$)$\times10^{-11}$ \\
$N$(H$_2$) [cm$^{-2}$] & ($7.7-9.5$)$\times10^{22}$ & ($6.7-11.8$)$\times10^{22}$ \\
Region mass [M$_{\odot}$] &  14.5 -- 17.9 & 9.3 -- 16.2 \\
\enddata
\end{deluxetable}

\subsection{How long will mass accretion last via the streamer?} \label{sec:dis3}

To evaluate the lifetime of the streamer, we first calculate the streamer infall rate ($\dot{M}_{\rm {streamer}}$).
There is a possibility that Region P2 contains gas outside of the streamer which will not eventually flow into the protostellar system.
To investigate this possibility, we calculated the mass of the streamer using the following formulae:
\begin{equation} \label{equ:mass2}
M_{\rm {new}} = M_{\rm {Pineda}} \frac{X_{\rm {Pineda}}}{X_{\rm {P2}}}.
\end{equation}
$M_{\rm {Pineda}}$ (= 0.1 M$_{\odot}$\,yr$^{-1}$) is the mass of the streamer derived by \citet{2020NatAs...4.1158P}.
$X_{\rm {Pineda}}$ ($=3.95\times10^{-9}$, which was converted into the value concerning the total hydrogen nuclei) is the HC$_3$N abundance assumed in \citet{2020NatAs...4.1158P}, and $X_{\rm {P2}}$ (=($3-7$)$\times10^{-11}$) means the HC$_3$N abundance, $X$(HC$_3$N), at P2 (Table \ref{tab:modelcomp}).
Using this formula, we obtained 5.6 -- 13.2 M$_{\odot}$ for the mass of the streamer.
Hence, the gas mass within Region P2 (9.3 -- 16.2 M$_{\odot}$) is higher than the newly calculated value by a factor of $\sim 1.4$.
This means that the single-dish observations trace gas from mostly the streamer, but small contamination from the extended gas is present.

To derive the streamer infall rate, we thus used the following formulae utilizing the result of \citet{2020NatAs...4.1158P} which only covers the streamer:
\begin{equation} \label{equ:infallrate}
\dot{M}_{\rm {streamer}} = \dot{M}_{\rm {Pineda}} \frac{X_{\rm {Pineda}}}{X_{\rm {P2}}},
\end{equation}
where $\dot{M}_{\rm {Pineda}}$ (= $10^{-6}$ M$_{\odot}$\,yr$^{-1}$) is the streamer infall rate derived by \citet{2020NatAs...4.1158P}.
The newly calculated streamer infall rate is ($5.6-13.2$)$\times10^{-5}$ M$_{\odot}$\,yr$^{-1}$. 

With the newly obtained streamer infall rate, we calculated the lifetime of the streamer ($t_{\rm {L}}$).
We assumed that all of the gas in Region P1 would continue to accrete onto the YSO with the current streamer infall rate and would not go to the ambient components.
We estimated the lifetime of the streamer with the following formulae:
\begin{equation} \label{equ:lifetime}
t_{\rm {L}} = \frac{M_{\rm {Region 1}}}{\dot{M}_{\rm {streamer}}},
\end{equation}
where $M_{\rm {Region 1}}$ corresponds to the mass within Region P1 (14.5 -- 17.9 M$_{\odot}$).
With the above assumption, the derived lifetime is (1.1 -- 3.2)$\times10^{5}$ yr, corresponding to a factor of 1.1 -- 3.3 of the free-fall timescale \citep[96 kyr;][]{2020NatAs...4.1158P}. 
Considering the typical evolution of YSOs \citep{2011AREPS..39..351D, 2014prpl.conf..195D}, mass accretion via the streamer will be able to continue until the end of the Class I stage.

Region P1 (the reservoir) is likely a starless core, but it may not be able to form a new protostar, because the material within this region is flowing into the Per-emb-2 protostellar system.
Note that Region P1 would chemically evolve, and the best tracers of the streamer may change with time.

A possible reason why the YSO Per-emb-2 is deficient in COMs \citep{2021ApJ...910...20Y} may be that the chemically young gas has been brought from another starless-core-like region and changed the chemical composition near the protostellar core.
To confirm this, we need high-angular-resolution observations to investigate chemical differentiation among carbon-chain species and CH$_3$OH around the YSO.

The P3 position may be an extra reservoir of material that contributes to the mass accretion into the YSO Per-emb-2, although we cannot measure an accretion rate from P3 due to the low-angular-resolution data.
The P3 position has the almost same centroid velocities as P1 (see Figure \ref{fig:HC3Nregrid} and Table \ref{tab:HC3Nline}), suggesting that they trace the same cloud material around P2, while P2 shows the red-shifted components that is the line-of-sight motion of material from the cloud toward the protostar.
The line width at P2 is more similar to that at P3 (Table \ref{tab:HC3Nline}), which suggests that accretion of material from P3 may currently be, or has been in the past, dominant.
Another possibility for the wide line widths at P2 is that the turbulence is caused by the protostellar activity.
We need high-angular-resolution observations covering P2 and P3 to clarify contributions from P3.

\section{Conclusions} \label{sec:con}

We conduct mapping observations of four carbon-chain species (HC$_3$N, HC$_5$N, CCH, and CCS) in the 3mm and 7mm bands toward the Per-emb-2 streamer in the Perseus region with the Nobeyama 45m radio telescope.
The main conclusions of this paper are as follows.

\begin{enumerate}

\item We find a structure at the north part of the protostellar system with a distance of $\sim 20,500$ au in integrated-intensity maps of the carbon-chain species.
This structure connects to the streamer and is identified as the reservoir of the streamer.

\item Combining the data obtained by the IRAM 30m telescope, we conduct the spectral-line energy distribution (SLED) analysis with the non-LTE RADEX code and the HC$_3$N lines.
We derive the density at the HC$_3$N emission peak within the reservoir (P1 position) $n_{\rm {H_{2}}}=$($1.9\pm0.3$)$\times10^4$ cm$^{-3}$ with the gas kinetic temperature of 10 K.
The position where the streamer connects to the YSO (P2 position) shows a slightly higher temperature and density; 14.5 K and ($5.1\pm0.6$)$\times10^4$ cm$^{-3}$.
The streamer system has likely almost homogeneous density and temperature structures.

\item We compare the observed CCS/HC$_3$N abundance ratios to the chemical model and constrain the relative abundances of HC$_3$N to the hydrogen nuclei. 
P1 and P2 are chemically young ($t \approx 10^4$ yr). 
This agrees with the fact that the Per-emb-2 streamer is well traced by carbon-chain species, whereas the emission of N$_2$H$^+$ is not consistent with the streamer structure \citep{2020NatAs...4.1158P}.
Moreover, similar chemical compositions at P1 and P2 strengthen our conclusion that Region P1 is the reservoir of the streamer.

\item The derived masses of the reservoir and the streamer are $\sim14.5-18$ M$_{\odot}$ and $\sim9-16$ M$_{\odot}$, respectively. Thus, the total mass available for the streamer is determined at 24 -- 34 M$_{\odot}$.

\item The streamer infall rate is recalculated at ($5.6 - 13.2$)$\times10^{-5}$ M$_{\odot}$\,yr$^{-1}$. We estimate the lifetime of the streamer assuming that all of the gas in the reservoir will flow into the Per-emb-2 protostellar system. The lifetime is calculated to be (1.1 -- 3.2)$\times10^{5}$ yr. The value exceeds the free-fall timescale by a factor of 1.1 -- 3.3.
The results suggest that the mass accretion via the streamer can continue until the end of the Class I phase.
\end{enumerate}

Streamers have been found by observations with interferometries toward YSOs at various evolutionary stages.
We demonstrate that single-dish mapping observations are also important to find out the origin(s) of streamers and derive their total masses. 


\begin{acknowledgments}
We would like to express thanks to the staff of the Nobeyama Radio Observatory.
The Nobeyama 45m radio telescope is operated by Nobeyama Radio Observatory, a branch of National Astronomical Observatory of Japan.
K.T. is supported by JSPS KAKENHI grant No. JP20K14523 and 21H01142. 
K.T. is grateful for research funds for NAOJ Fellow to purchase the observing time of the Nobeyama 45m radio telescope and to support with travel funding for visiting Max-Planck-Institut f\"{u}r Extraterrestrische Physik (MPE).
K.T. greatly appreciates NAOJ Overseas Visit Program for Young Researchers 2023 for support with travel funding for visiting MPE. 
T.S. has been financially supported by JSPS KAKENHI grant No. 22K02966 and the Kayamori Foundation of Informational Science Advancement. 
DMS is supported by an NSF Astronomy and Astrophysics Postdoctoral Fellowship under award AST-2102405.
We appreciate an anonymous referee whose comments are helpful to improve the paper.
\end{acknowledgments}

\vspace{5mm}
\facilities{Nobeyama 45m radio telescope}

\software{Common Astronomy Software Applications package \citep[CASA;][]{2022PASP..134k4501C}, CASSIS \citep{2015sf2a.conf..313V}, Nautilius \citep{2016MNRAS.459.3756R}}

\appendix

\section{Channel maps of carbon-chain species in the 3mm band} \label{sec:a0}

\begin{figure*}
 \begin{center}
  \includegraphics[bb = 0 10 470 555, scale =0.85]{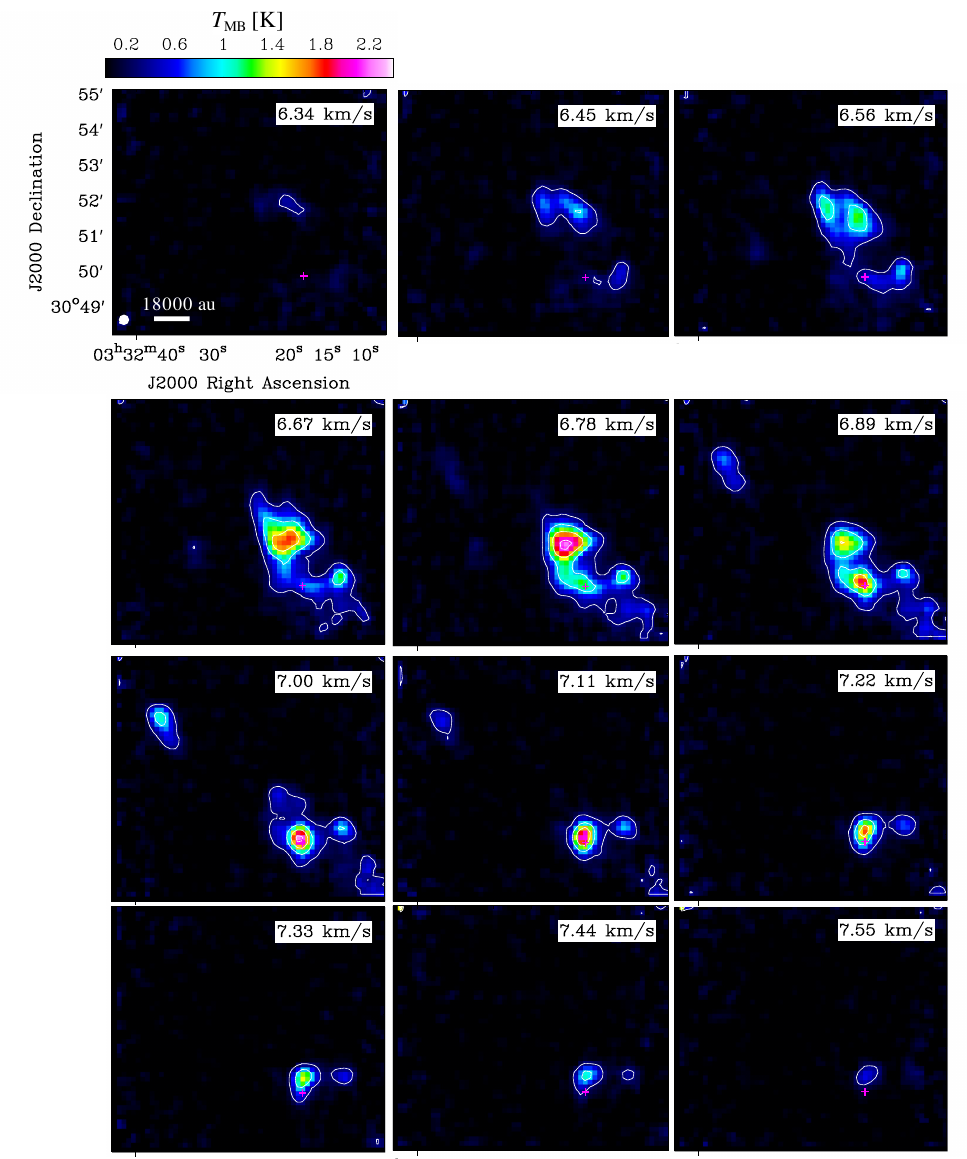}
 \end{center}
\caption{Channel maps of the HC$_3$N ($J=9-8$) line. The magenta cross indicates the position of the YSO. The contour levels are 10, 30, 50, 70$\sigma$, where $\sigma = 0.03$ K. The white-filled circle at the bottom left corner in the first panel indicates the beam size at 80 GHz (18\arcsec). \label{fig:HC3Nchan}}
\end{figure*}

\begin{figure*}
 \begin{center}
  \includegraphics[bb = 0 10 470 510, scale =0.92]{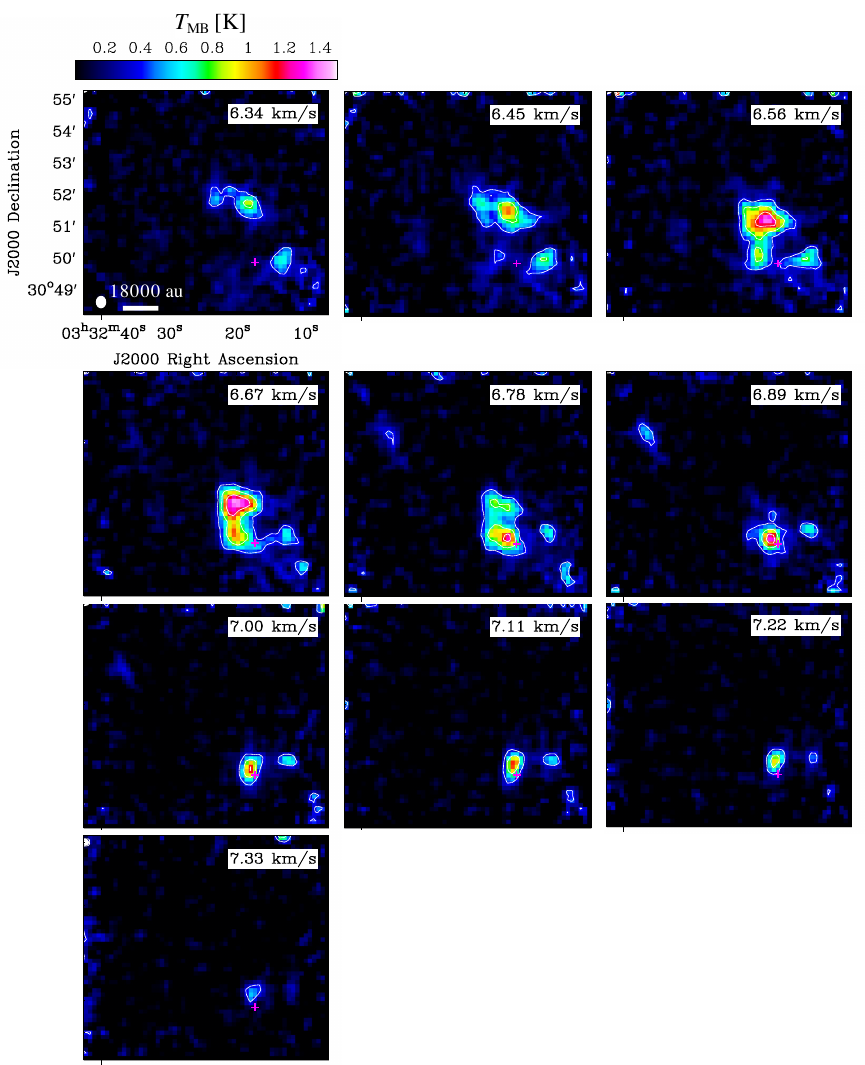}
 \end{center}
\caption{Channel maps of the CCS ($J_N=6_7-5_6$) line. The magenta cross indicates the position of the YSO. The contour levels are 10, 20, 30$\sigma$, where $\sigma = 0.038$ K. The white-filled circle at the bottom left corner in the first panel indicates the beam size at 80 GHz (18\arcsec). \label{fig:CCSchan}}
\end{figure*}

\begin{figure*}
 \begin{center}
  \includegraphics[bb = 0 10 470 590, scale =0.88]{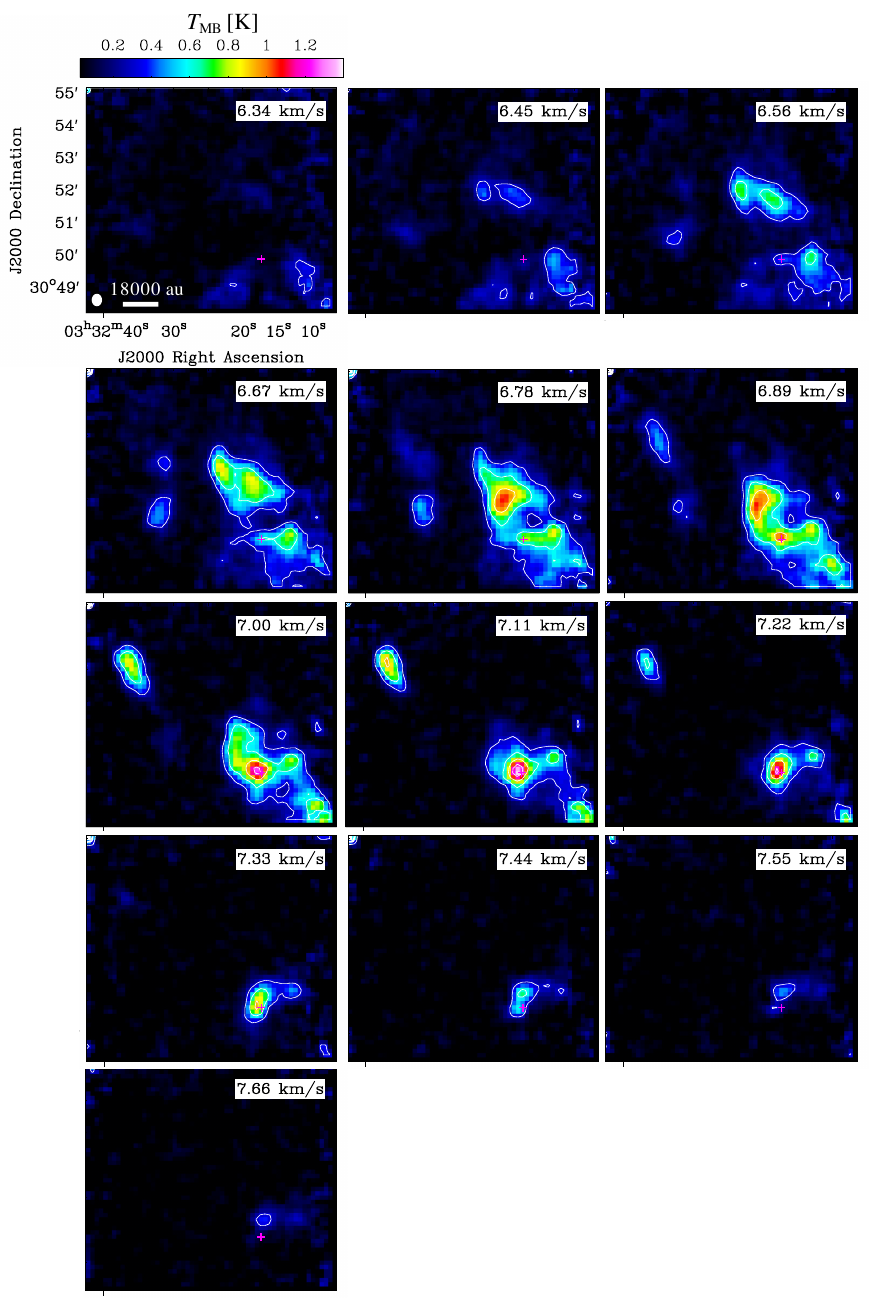}
 \end{center}
\caption{Channel maps of the CCH ($N= 1- 0$, $J=\frac{3}{2}-\frac{1}{2}$, $F= 2- 1$) line. The magenta cross indicates the position of the YSO. The contour levels are 10, 20, 30, 40$\sigma$, where $\sigma = 0.03$ K. The white-filled circle at the bottom left corner in the first panel indicates the beam size at 80 GHz (18\arcsec). \label{fig:CCHchan}}
\end{figure*}

Figures \ref{fig:HC3Nchan}, \ref{fig:CCSchan}, and \ref{fig:CCHchan} show channel maps of the HC$_3$N ($J=9-8$), CCS ($J_N=6_7-5_6$), and CCH ($N= 1- 0$, $J=\frac{3}{2}-\frac{1}{2}$, $F= 2- 1$) lines, respectively.
Velocity gradients from the northern part to the YSO can be seen in these channel maps.
All of the three molecular lines show similar structures.
The upper-state energy of the CCH line is lower than those of the other two species (Table \ref{tab:line}).
This explains the more extended morphology of CCH (1--0) compared to the other lines. 

\section{Gaussian fitting results of HC$_3$N and CCS lines} \label{sec:a1}

Table \ref{tab:HC3Nline} summarizes line parameters obtained by the Gaussian fitting of the HC$_3$N ($J=9-8$) line at P1, P2, and P3.
The right panel of Figure \ref{fig:HC3Nregrid} shows the observed spectra overlaid by these Gaussian fitting results.

\begin{deluxetable}{lccc}
\tablecaption{Line parameters of the HC$_3$N ($J=9-8$) line \label{tab:HC3Nline}}
\tablewidth{0pt}
\tablehead{
\colhead{Position} & \colhead{$T_{\rm {MB}}$ (K)} & \colhead{$V_{\rm{LSR}}$ (km\,s$^{-1}$)} & \colhead{FWHM (km\,s$^{-1}$)} 
}
\startdata
P1 & $2.47 \pm 0.02 $ & $6.751 \pm 0.001$ & $0.331 \pm 0.003$ \\    
P2 & $1.77 \pm 0.02$ & $7.009 \pm 0.004$ & $0.65 \pm 0.01$ \\
P3 & $0.93 \pm 0.03$ & $6.82 \pm 0.01$ & $0.69 \pm 0.03$ \\
\enddata
\tablecomments{The errors indicate the standard deviation.} 
\end{deluxetable}

\begin{deluxetable*}{lccccccc}
\tablecaption{Line parameters of the CCS lines \label{tab:CCSline}}
\tablewidth{0pt}
\tablehead{
\colhead{} & \multicolumn{3}{c}{P1} & & \multicolumn{3}{c}{P2} \\
\cline{2-4} \cline{6-8}
\colhead{Transition} & \colhead{$T_{\rm {MB}}$ (K)} & \colhead{$V_{\rm{LSR}}$ (km\,s$^{-1}$)} & \colhead{FWHM (km\,s$^{-1}$)} & \colhead{} &  \colhead{$T_{\rm {MB}}$ (K)} & \colhead{$V_{\rm{LSR}}$ (km\,s$^{-1}$)} & \colhead{FWHM (km\,s$^{-1}$)}
}
\startdata
$4_3-3_2$ & $1.74 \pm 0.06$ & $6.858 \pm 0.005$ & $0.322 \pm 0.012$ & & $0.86 \pm 0.05$ & $7.095 \pm0.018$ & $0.60 \pm 0.04$ \\
$6_7-5_6$ & $1.15 \pm 0.02$ & $6.631 \pm 0.003$ & $0.311 \pm 0.007$ & & $0.813 \pm 0.019$ & $6.844 \pm 0.007$ & $0.645 \pm 0.017$ \\ 
$2_1-1_2$ & $1.16 \pm 0.12$ & $6.750 \pm 0.012$ & $0.24 \pm 0.03$ & & $0.39 \pm 0.06$ & $7.06 \pm 0.05$ & $0.68 \pm 0.13$ \\
\enddata
\tablecomments{The errors indicate the standard deviation.} 
\end{deluxetable*}

Table \ref{tab:CCSline} summarizes line parameters of the CCS lines at P1 and P2 obtained by fitting with the Gaussian profile.
The spectra with the results of the Gaussian fitting are shown in the upper panels of Figure \ref{fig:RD}.

\section{Comparisons of the CCS/HC$_5$N abundance ratios} \label{sec:a2}

\begin{figure}
 \begin{center}
  \includegraphics[bb = 0 30 680 510, scale = 0.5]{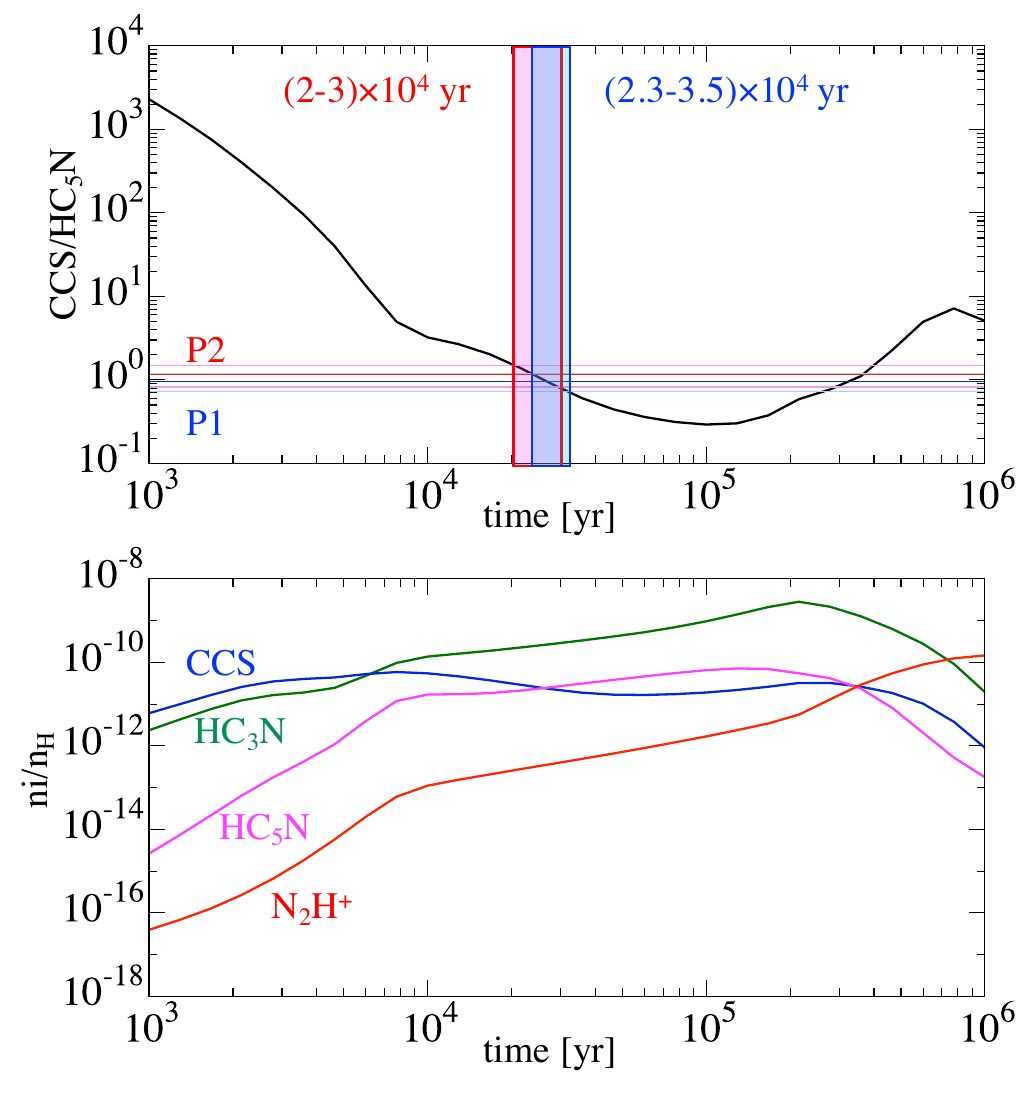}
 \end{center}
\caption{The upper panel compares the observed abundance ratios of CCS/HC$_5$N at P1 and P2 to the modeled ones. The horizontal lines show the observed values (light-colored lines indicate the upper and lower limits). The lower panels show the abundances of three carbon-chain species and N$_2$H$^+$. \label{fig:modelHC5N}}
\end{figure}

The upper panel of Figure \ref{fig:modelHC5N} shows the comparisons between the observed CCS/HC$_5$N ratios at P1 and P2 to the modeled one.
The observed ratios are $0.98\pm0.23$ and $1.16\pm0.34$ at P1 and P2, respectively.
Their relative abundances agree with each other within their uncertainties.
The best-fit ages at P1 and P2 are almost the same age of (2.3 -- 3.0)$\times10^4$ yr.
This is close to the best agreement age obtained from the comparisons of CCS/HC$_3$N ratios.

The lower panel of Figure \ref{fig:modelHC5N} shows the time evolution of molecular abundances of the three carbon-chain species and N$_2$H$^+$.
The abundance of N$_2$H$^+$ around $10^4$ yr is below $10^{-13}$, which is lower than those of carbon-chain species by more than two orders of magnitude.

\begin{thebibliography}{}
\bibitem[Alves et al.(2019)]{2019Sci...366...90A} Alves, F.~O., Caselli, P., Girart, J.~M., et al.\ 2019, Science, 366, 90. doi:10.1126/science.aaw3491
\bibitem[Arzoumanian et al.(2023)]{2023ApJ...947L..29A} Arzoumanian, D., Arakawa, S., Kobayashi, M.~I.~N., et al.\ 2023, \apjl, 947, L29. doi:10.3847/2041-8213/acc849
\bibitem[Benson et al.(1998)]{1998ApJ...506..743B} Benson, P.~J., Caselli, P., \& Myers, P.~C.\ 1998, \apj, 506, 743. doi:10.1086/306276
\bibitem[Bate(2018)]{2018MNRAS.475.5618B} Bate, M.~R.\ 2018, \mnras, 475, 5618. doi:10.1093/mnras/sty169
\bibitem[CASA Team et al.(2022)]{2022PASP..134k4501C} CASA Team, Bean, B., Bhatnagar, S., et al.\ 2022, \pasp, 134, 114501. doi:10.1088/1538-3873/ac9642
\bibitem[Chou et al.(2016)]{2016ApJ...823..151C} Chou, H.-G., Yen, H.-W., Koch, P.~M., et al.\ 2016, \apj, 823, 151. doi:10.3847/0004-637X/823/2/151
\bibitem[Cuello et al.(2020)]{2020MNRAS.491..504C} Cuello, N., Louvet, F., Mentiplay, D., et al.\ 2020, \mnras, 491, 504. doi:10.1093/mnras/stz2938
\bibitem[Dauphas \& Chaussidon(2011)]{2011AREPS..39..351D} Dauphas, N. \& Chaussidon, M.\ 2011, Annual Review of Earth and Planetary Sciences, 39, 351. doi:10.1146/annurev-earth-040610-133428 
\bibitem[Dobashi et al.(2018)]{2018ApJ...864...82D} Dobashi, K., Shimoikura, T., Nakamura, F., et al.\ 2018, \apj, 864, 82. doi:10.3847/1538-4357/aad62f
\bibitem[Dunham et al.(2014)]{2014prpl.conf..195D} Dunham, M.~M., Stutz, A.~M., Allen, L.~E., et al.\ 2014, Protostars and Planets VI, 195. doi:10.2458/azu\_uapress\_9780816531240-ch009
\bibitem[Endres et al.(2016)]{2016JMoSp.327...95E} Endres, C.~P., Schlemmer, S., Schilke, P., et al.\ 2016, Journal of Molecular Spectroscopy, 327, 95. doi:10.1016/j.jms.2016.03.005
\bibitem[Faure et al.(2016)]{2016MNRAS.460.2103F} Faure, A., Lique, F., \& Wiesenfeld, L.\ 2016, \mnras, 460, 2103. doi:10.1093/mnras/stw1156
\bibitem[Fern{\'a}ndez-L{\'o}pez et al.(2023)]{2023ApJ...956...82F} Fern{\'a}ndez-L{\'o}pez, M., Girart, J.~M., L{\'o}pez-V{\'a}zquez, J.~A., et al.\ 2023, \apj, 956, 82. doi:10.3847/1538-4357/ace786
\bibitem[Friesen et al.(2017)]{2017ApJ...843...63F} Friesen, R.~K., Pineda, J.~E., co-PIs, et al.\ 2017, \apj, 843, 63. doi:10.3847/1538-4357/aa6d58
\bibitem[Garufi et al.(2022)]{2022A&A...658A.104G} Garufi, A., Podio, L., Codella, C., et al.\ 2022, \aap, 658, A104. doi:10.1051/0004-6361/202141264
\bibitem[Ginski et al.(2021)]{2021ApJ...908L..25G} Ginski, C., Facchini, S., Huang, J., et al.\ 2021, \apjl, 908, L25. doi:10.3847/2041-8213/abdf57
\bibitem[Hirota et al.(2009)]{2009ApJ...699..585H} Hirota, T., Ohishi, M., \& Yamamoto, S.\ 2009, \apj, 699, 585. doi:10.1088/0004-637X/699/1/585
\bibitem[Hsieh et al.(2023)]{2023A&A...669A.137H} Hsieh, T.-H., Segura-Cox, D.~M., Pineda, J.~E., et al.\ 2023, \aap, 669, A137. doi:10.1051/0004-6361/202244183
\bibitem[Kaifu et al.(2004)]{2004PASJ...56...69K} Kaifu, N., Ohishi, M., Kawaguchi, K., et al.\ 2004, \pasj, 56, 69. doi:10.1093/pasj/56.1.69
\bibitem[Kamazaki et al.(2012)]{2012PASJ...64...29K} Kamazaki, T., Okumura, S.~K., Chikada, Y., et al.\ 2012, \pasj, 64, 29. doi:10.1093/pasj/64.2.29
\bibitem[Le Gouellec et al.(2019)]{2019ApJ...885..106L} Le Gouellec, V.~J.~M., Hull, C.~L.~H., Maury, A.~J., et al.\ 2019, \apj, 885, 106. doi:10.3847/1538-4357/ab43c2
\bibitem[Lee et al.(2023)]{2023ApJ...953...82L} Lee, J.-E., Matsumoto, T., Kim, H.-J., et al.\ 2023, \apj, 953, 82. doi:10.3847/1538-4357/acdd5b
\bibitem[Mangum et al.(2007)]{2007A&A...474..679M} Mangum, J.~G., Emerson, D.~T., \& Greisen, E.~W.\ 2007, \aap, 474, 679. doi:10.1051/0004-6361:20077811 
\bibitem[Mottram et al.(2017)]{2017A&A...600A..99M} Mottram, J.~C., van Dishoeck, E.~F., Kristensen, L.~E., et al.\ 2017, \aap, 600, A99. doi:10.1051/0004-6361/201628682
\bibitem[M{\"u}ller et al.(2005)]{2005JMoSt.742..215M} M{\"u}ller, H.~S.~P., Schl{\"o}der, F., Stutzki, J., et al.\ 2005, Journal of Molecular Structure, 742, 215. doi:10.1016/j.molstruc.2005.01.027
\bibitem[Minamidani et al.(2016)]{2016SPIE.9914E..1ZM} Minamidani, T., Nishimura, A., Miyamoto, Y., et al.\ 2016, \procspie, 9914, 99141Z. doi:10.1117/12.2232137
\bibitem[Murillo et al.(2022)]{2022A&A...658A..53M} Murillo, N.~M., van Dishoeck, E.~F., Hacar, A., et al.\ 2022, \aap, 658, A53. doi:10.1051/0004-6361/202141250
\bibitem[Nakamura et al.(2015)]{2015PASJ...67..117N} Nakamura, F., Ogawa, H., Yonekura, Y., et al.\ 2015, \pasj, 67, 117. doi:10.1093/pasj/psv088
\bibitem[Pickett et al.(1998)]{1998JQSRT..60..883P} Pickett, H.~M., Poynter, R.~L., Cohen, E.~A., et al.\ 1998, \jqsrt, 60, 883. doi:10.1016/S0022-4073(98)00091-0
\bibitem[Pineda et al.(2023)]{2023ASPC..534..233P} Pineda, J.~E., Arzoumanian, D., Andre, P., et al.\ 2023, Protostars and Planets VII, 534, 233. doi:10.48550/arXiv.2205.03935
\bibitem[Pineda et al.(2020)]{2020NatAs...4.1158P} Pineda, J.~E., Segura-Cox, D., Caselli, P., et al.\ 2020, Nature Astronomy, 4, 1158. doi:10.1038/s41550-020-1150-z
\bibitem[Podio et al.(2017)]{2017MNRAS.470L..16P} Podio, L., Codella, C., Lefloch, B., et al.\ 2017, \mnras, 470, L16. doi:10.1093/mnrasl/slx068
\bibitem[Pokhrel et al.(2018)]{2018ApJ...853....5P} Pokhrel, R., Myers, P.~C., Dunham, M.~M., et al.\ 2018, \apj, 853, 5. doi:10.3847/1538-4357/aaa240
\bibitem[Rodr{\'\i}guez-Fern{\'a}ndez et al.(2010)]{2010A&A...516A..98R} Rodr{\'\i}guez-Fern{\'a}ndez, N.~J., Tafalla, M., Gueth, F., et al.\ 2010, \aap, 516, A98. doi:10.1051/0004-6361/201013997
\bibitem[Ruaud et al.(2016)]{2016MNRAS.459.3756R} Ruaud, M., Wakelam, V., \& Hersant, F.\ 2016, \mnras, 459, 3756. doi:10.1093/mnras/stw887
\bibitem[Schw{\"o}rer et al.(2019)]{2019A&A...628A...6S} Schw{\"o}rer, A., S{\'a}nchez-Monge, {\'A}., Schilke, P., et al.\ 2019, \aap, 628, A6. doi:10.1051/0004-6361/201935200
\bibitem[Seifried et al.(2015)]{2015MNRAS.446.2776S} Seifried, D., Banerjee, R., Pudritz, R.~E., et al.\ 2015, \mnras, 446, 2776. doi:10.1093/mnras/stu2282
\bibitem[Stephens et al.(2019)]{2019ApJS..245...21S} Stephens, I.~W., Bourke, T.~L., Dunham, M.~M., et al.\ 2019, \apjs, 245, 21. doi:10.3847/1538-4365/ab5181
\bibitem[Suzuki et al.(1992)]{1992ApJ...392..551S} Suzuki, H., Yamamoto, S., Ohishi, M., et al.\ 1992, \apj, 392, 551. doi:10.1086/171456
\bibitem[Takakuwa et al.(2017)]{2017ApJ...837...86T} Takakuwa, S., Saigo, K., Matsumoto, T., et al.\ 2017, \apj, 837, 86. doi:10.3847/1538-4357/aa6116
\bibitem[Taniguchi et al.(2019)]{2019ApJ...884..167T} Taniguchi, K., Herbst, E., Ozeki, H., et al.\ 2019, \apj, 884, 167. doi:10.3847/1538-4357/ab3eb8
\bibitem[Taniguchi et al.(2016)]{2016ApJ...817..147T} Taniguchi, K., Ozeki, H., Saito, M., et al.\ 2016, \apj, 817, 147. doi:10.3847/0004-637X/817/2/147
\bibitem[Taniguchi et al.(2017)]{2017ApJ...844...68T} Taniguchi, K., Saito, M., Hirota, T., et al.\ 2017, \apj, 844, 68. doi:10.3847/1538-4357/aa7899

\bibitem[Valdivia-Mena et al.(2022)]{2022A&A...667A..12V} Valdivia-Mena, M.~T., Pineda, J.~E., Segura-Cox, D.~M., et al.\ 2022, \aap, 667, A12. doi:10.1051/0004-6361/202243310
\bibitem[van der Tak et al.(2007)]{2007A&A...468..627V} van der Tak, F.~F.~S., Black, J.~H., Sch{\"o}ier, F.~L., et al.\ 2007, \aap, 468, 627. doi:10.1051/0004-6361:20066820
\bibitem[Vastel et al.(2015)]{2015sf2a.conf..313V} Vastel, C., Bottinelli, S., Caux, E., et al.\ 2015, SF2A-2015: Proceedings of the Annual meeting of the French Society of Astronomy and Astrophysics, 313
\bibitem[Vorobyov et al.(2017)]{2017A&A...608A.107V} Vorobyov, E.~I., Steinrueck, M.~E., Elbakyan, V., et al.\ 2017, \aap, 608, A107. doi:10.1051/0004-6361/201731565
\bibitem[Yang et al.(2021)]{2021ApJ...910...20Y} Yang, Y.-L., Sakai, N., Zhang, Y., et al.\ 2021, \apj, 910, 20. doi:10.3847/1538-4357/abdfd6
\bibitem[Zucker et al.(2018)]{2018ApJ...869...83Z} Zucker, C., Schlafly, E.~F., Speagle, J.~S., et al.\ 2018, \apj, 869, 83. doi:10.3847/1538-4357/aae97c
\end{thebibliography}
\end{document}